\documentclass[prb,twocolumn,showpacs,preprintnumbers,float]{revtex4-2}
\usepackage[colorlinks=true,urlcolor=blue,citecolor=blue,linkcolor=blue,breaklinks=true]{hyperref}
\usepackage{graphicx}
\usepackage{amsmath}
\usepackage{amssymb}
\usepackage{times}
\usepackage{color}
\usepackage{ulem}
\usepackage{physics}
\usepackage{cleveref}
\usepackage{xr}
\usepackage{multirow}
\usepackage{comment}

\def\a{\alpha}

\def\D{\Delta}
\def\d{\delta}

\def\g{\gamma}

\def\l{\lambda}

\def\s{\sigma}

\def\t{\tau}
\def\w{\omega}

\def\cc{\hat{c}}

\def\Emax{E_{\rm max}}
\def\Etot{E_{\rm tot}}
\def\Aqsym{A_{q}^{\rm sym}}
\def\Bb{\bar{B}}
\def\Bbq{\bar{B}_{q}}
\def\Bbqsym{\bar{B}_{q}^{\rm sym}}
\newcommand{\nb}{\bar{n}}

\begin{document}

\title{Quantum phase transition between hyperuniform density distributions}

\author{Shiro Sakai$^1$, Ryotaro Arita$^{1,2}$, and Tomi Ohtsuki$^3$}
\affiliation{
$^1$Center for Emergent Matter Science, RIKEN, Wako, Saitama 351-0198, Japan\\
$^2$Research Center for Advanced Science and Technology, University of Tokyo, Komaba, Tokyo 153-8904, Japan\\
$^3$Physics Division, Sophia University, Chiyoda-ku, Tokyo 102-8554, Japan
}
\date{\today}
\begin{abstract}
We study an electron distribution under a quasiperiodic potential in light of hyperuniformity, aiming to establish a classification and analysis method for aperiodic but orderly density distributions realized in, e.g., quasicrystals.
Using the Aubry-Andr\'e-Harper model, we first reveal that the electron-charge distribution changes its character as the increased quasiperiodic potential alters the eigenstates from extended to localized ones.
While these changes of the charge distribution are characterized by neither multifractality nor translational-symmetry breaking, they are characterized by hyperuniformity class and its order metric.
We find a nontrivial relationship between the density of states at the Fermi level, a charge-distribution histogram, and the hyperuniformity class. 
The change to a different hyperuniformity class occurs as a first-order phase transition except for an electron-hole symmetric point, where the transition is of the third order.
Moreover, we generalize the hyperuniformity order metric to a function, to capture more detailed features of the density distribution, in some analogy with a generalization of the fractal dimension to a multifractal one.
\end{abstract}
\maketitle

\section{introduction}\label{sec:intro}
Inhomogeneous but orderly electron states, realized in quasicrystals \cite{mackay82,shechtman84,levine84}, possess properties distinct from both periodic and random systems.
Early studies of tight-binding Fibonacci models \cite{kohmoto87,sutherland87,tokihiro88,mace17,jagannathan20} showed that the density of states (DOS) is singular continuous and that the eigenstates are multifractal \cite{halsey86}.
However, quasiperiodic electron states are not always multifractal.
For instance, the Aubry-Andr\'e-Harper (AAH) model \cite{aubry80,harper55}, which has a quasiperiodic potential incommensurate to the lattice periodicity, shows a multifractality only at a special strength of the potential. 
Moreover,  besides eigenfunctions, we can consider spatial distributions of various electron properties like the electron density \cite{sakai21,sakai22}, magnetization in quasiperiodic magnets \cite{wessel03,vedmedenko04,wessel05,jagannathan07,thiem15,koga17,koga20,tamura21,watanabe21}, and order parameter in quasiperiodic superconductors \cite{sakai17,kamiya18,araujo19,sakai19,nagai20,takemori20}.
These distributions are not necessarily multifractal while they still show interesting orderly but aperiodic patterns \cite{sakai22}.

In Ref.~\onlinecite{sakai22}, we showed that the electron-charge distribution on the Penrose tiling, as well as of the AAH model, is characterized by hyperuniformity.
Hyperuniformity, coined by Torquato and his collaborators \cite{torquato03,torquato18}, is a framework to quantify the regularity of the spatial distribution of a point set and has been generalized to a random scalar field \cite{torquato16,ma17,torquato18}.
It measures a density fluctuation of a given point set or scalar field distributed in a $d$-dimensional space and distinguishes different distributions according to the strength of the density fluctuation at a large length scale.
Periodic and quasiperiodic point sets (i.e., lattice) are known to be hyperuniform. Namely, they possess significantly small density fluctuations thanks to the regularity of the lattices.
 

Various quasiperiodic lattices (as point sets) have then been classified in terms of hyperuniformity classes and its order metric \cite{torquato18,oguz17,lin18}, which quantify the degree of regularity of a hyperuniform distribution.
The relevance of the order metric to a band-gap size of photonic quasicrystals has also been suggested \cite{florescu09}.
In contrast, the nature of hyperuniform electron states or distributions (as scalar fields) realized on quasiperiodic structures remains largely unexplored.
In particular, unlike periodic systems, where the change of the charge distribution occurs as a phase transition accompanied by the translational-symmetry breaking, it is unclear if such a change on quasiperiodic lattices occurs as a phase transition since the translational symmetry is absent in the first place.

In this paper, we scrutinize the AAH model \cite{aubry80,harper55,Sokoloff85}, which is a prototypical quasiperiodic model in one dimension and has been realized experimentally in ultracold atoms \cite{roati08} and photonic quasicrystals \cite{lahini09}, in light of the hyperuniformity.
Because the AAH model exhibits extended, critical, and localized eigenstates according to the strength of the quasiperiodic potential \footnote{This is distinct from the Fibonacci model, where the eigenstates are always critical. We study the Fibonacci model in Appendix \ref{sec:fib}.}, it has long been studied in the context of quantum localization.
Here, we utilize this property to study the relationship between the electron-localization strength and hyperuniformity, focusing on the charge distributions rather than the eigenstates. 

We find that the charge distribution in the AAH model is always hyperuniform but its class and the order metric change according to the quasiperiodic potential and the Fermi level.
For a weak potential, where the eigenstates are extended, the charge distribution has no jump in its histogram, exhibiting Class-I hyperuniform behavior.
At and above the self-dual point \cite{aubry80,Sokoloff85}, where the eigenstates are critical and localized, respectively, the charge distribution has no jump and Class-I hyperuniform only when the DOS at the Fermi level vanishes; otherwise, it has a vanishing point or a jump in the histogram and belongs to Class II. 
We thus reveal a nontrivial relationship between the DOS, charge distribution, and hyperuniformity class.
We then clarify that the change of the hyperuniformity class is the first-order phase transition except for the electron-hole symmetric point where it is of the third order.
These results, in turn, uncover a significant difference of the AAH model from random systems, where the localized states do not constitute a hyperuniform charge distribution.

Furthermore, we generalize the order metric to a function for a hyperuniform scalar-field distribution, in some analogy with the generalization of the fractal dimension to the multifractal one \cite{halsey86}.
This generalization allows us to quantify more detailed features of density distributions.
This ``multihyperuniformity'' \footnote{The term ``multihyperuniformity'' has been used in Refs.\cite{jiao14,lomba20} for point patterns that their multiple distinct subsets are hyperuniform. In this paper, we define ``multihyperuniformity'' for density distributions.} would be useful to characterize various density distributions, which are not multifractal but hyperuniform.

The rest of the paper is organized as follows. 
In Sec.~\ref{sec:method}, we introduce the AAH Hamiltonian and the method to calculate its electron distribution and hyperuniformity.
In Secs.~\ref{ssec:dos} and \ref{ssec:ni}, we show the results of the DOS and charge distribution for various strengths of the quasiperiodic potential, revealing a relation between them.
In Sec.~\ref{ssec:hyper_results}, we discuss the results of hyperuniformity for the charge distributions and find that its class changes with the DOS at the Fermi energy as well as the continuity of the charge-distribution histogram.
In Sec.~\ref{ssec:phase}, we reveal that the abrupt change of the hyperuniformty is indeed a phase transition.
In Sec.~\ref{ssec:multi}, we introduce the ``multi-hyperuniformity" to characterize more details of the density distribution.
Section \ref{sec:summary} summarizes the paper. 
In Appendix \ref{sec:fib}, we compare the results with those obtained for the Fibonacci models, to find a similarity to the critical case of the AAH model.
Appendix \ref{sec:random} is devoted to demonstrate that the charge distribution under a random potential is not hyperuniform.
Appendix \ref{sec:zk} shows the results for an integrated intensity function, which gives an alternative way to calculate the hyperuniformity class.
In Appendix \ref{sec:class2}, we demonstrate that Class-II hyperuniform distributions remain Class II in our definition of the ``multi-hyperuniformity".
Appendix \ref{sec:n2} presents the results of a local variance.


\section{Model and Method}\label{sec:method}
\subsection{Aubry-Andr\'e-Harper model}\label{ssec:model}

\begin{figure*}[tb]
\center{
\includegraphics[width=0.9\textwidth]{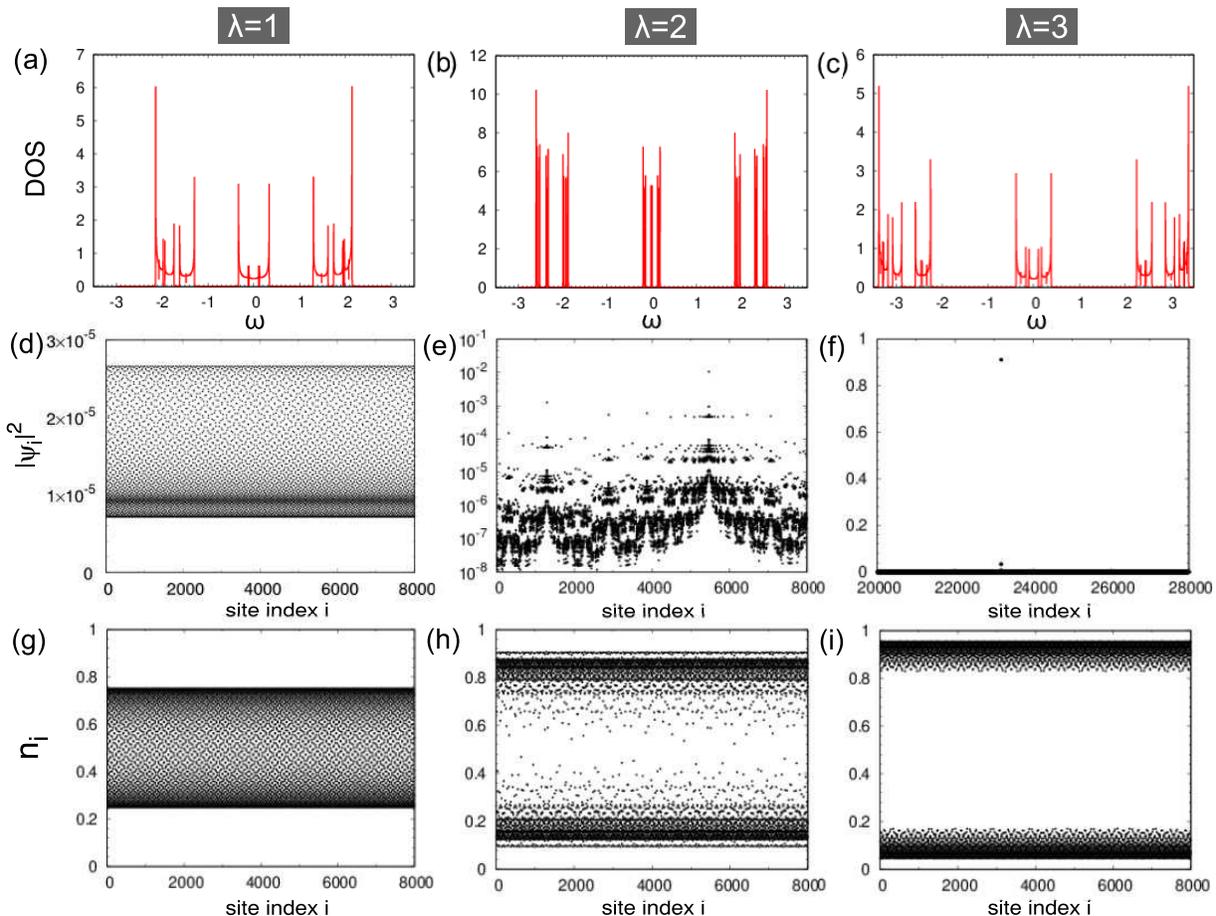}}
\caption{(a,b,c) Density of states, (d,e,f) the amplitude of the lowest-energy eigenfunction, and (g,h,i) the charge distribution for $\l=1$, 2, and 3, respectively, at $\mu=0$.}
\label{fig:aah}
\end{figure*}

The AAH Hamiltonian \cite{aubry80,harper55} reads
\begin{align}
 H&=-t\sum_i \left(e^{-i\phi}\cc_{i+1}^\dagger \cc_i+{\rm h.c.}\right)\nonumber\\
   &+\sum_i \left[\lambda \cos(\frac{2\pi i}{\t}+\phi)-\mu\right]\cc_{i}^\dagger \cc_i,\label{eq:aah}
\end{align}
where $\hat{c}_{i}$ ($\cc_{i}^\dagger$) annihilates (creates) a spinless fermion (which we call electron in this paper) at a site $i$ on a one-dimensional chain with the lattice constant $a=1$.
$t$ represents the hopping integral to the neighboring sites and $\lambda$ does the strength of the quasiperiodic potential, where 
$\t=\frac{\sqrt{5}+1}{2}$ is the golden ratio. We set $t=1$ and use it as the unit of energy. 
We have added the chemical potential ($\mu$) term to the original Hamiltonian, to discuss the relation between the spectrum and the charge distribution.
Considering zero temperature, we define $\nb\equiv \frac{1}{N}\sum_{i} n_{i}$ with the number of sites $N$ and 
\begin{align}
  n_i &\equiv\langle \cc_{i}^\dagger \cc_{i}\rangle =\sum_{E_\a<0} \langle \psi_\a | \cc_i^\dagger \cc_i |\psi_\a \rangle, \label{eq:ni}
\end{align}
where $\psi_\a$ and $E_\a$ are the eigenstates and eigenenergies of the Hamiltonian (\ref{eq:aah}), respectively. 

This model is known to be self-dual at $\lambda=2t$. 
Namely, the form of the Hamiltonian does not change after the Fourier transformation to momentum space with exchanging $\l$ and $2t$.
As a consequence, the eigenfunctions are extended (localized) in real space for $\lambda<2t$ ($\lambda>2t$) and critical at $\lambda=2t$ [see Figs.~\ref{fig:aah}(d,e,f)].

We numerically diagonalize a one-dimensional chain of $N=F_n$ sites, where $F_n$ is the $n$-th Fibonacci number.
We use $N=F_{24}=75025$ with a periodic boundary condition, where $\t$ in Eq.~(\ref{eq:aah}) is approximated by $\frac{F_n}{F_{n-1}}$. 
In this case, the phase shift $\phi$ does not play a significant role in the eigenvalues, unlike the topological surface states observed for the open-boundary condition \cite{kraus12}. We therefore set $\phi=0$ in the following.
By comparing the results with those obtained with other sizes, we have confirmed that $N=F_{24}$ is sufficiently large to infer the infinite-size limit.

\subsection{Hyperuniformity}\label{ssec:hyper}
Hyperuniformity is a framework to distinguish and quantify various spatial distributions.
It was invented by Torquato and Stillinger \cite{torquato03} originally for point patterns distributed in space and has been generalized to various types of distribution including a random scalar field \cite{torquato16,torquato18,ma17}.

In one dimension, we consider a window of a range $[-R, R]$ and count the number of points (or the total value of the scalar field) contained in the window.
Namely, denoting the center position of the window as $r_c$, we calculate the quantity,
\begin{align}
 N(R)=\sum_{i} n_i \Theta(R-|r_i-r_c|) \label{eq:nr}
 \end{align}
 with the Heaviside step function $\Theta(r)$.
Then, its variance is given by
\begin{align}
  \s^2(R) = \overline{N(R)^2} - \left[ \overline{N(R)} \right]^2, \label{eq:s2}
\end{align}
where $\overline{Q}$ represents the average of $Q$ with respect to the center position $r_c$ over the system.
While $\s^2(R)$ is proportional to $R^d$ (with $d=1$ in the present case) for a random distribution of $n_i$, the distribution with $\s^2(R)<O(R^d)$ is called hyperuniform, which means that a bulk contribution to the variance vanishes.
Hyperuniform distributions are further classified into several classes: 
In one dimension, a distribution is called Class-I and Class-II hyperuniform, respectively, when the large-$R$ behavior of $\s^2(R)$ is constant and proportional to $\log R$ \cite{torquato18}.
Point distributions (i.e., $n_i\equiv 1$) on periodic and quasiperiodic lattices are known to be hyperuniform \cite{torquato03,torquato18}.

To judge if a one-dimensional distribution is hyperuniform from a finite-size calculation, we consider the following function,
\begin{align}
    A(R)=\s^2(R)/R.\label{eq:ar}
\end{align}
If $A(R)$ goes to zero as $R$ increases, the distribution is hyperuniform.
In particular, when it is Class I, i.e., $\s^2(R)= {\rm const.}$ for a large $R$, we define \cite{torquato18}
\begin{align}
  \Bb(R) \equiv\frac{1}{\nb^2 R} \int_0^R \s^2(R') dR'.  \label{eq:br}
\end{align} 
Namely, we average over $[0, R]$ to infer the order metric $\Bb(\infty)$ because $\s^2(R)$ typically oscillates with $R$ around its mean value.
The factor $1/\nb^2$ is just to eliminate a trivial contribution from $\nb$ to $\s^2(R)$.


\section{Results and Discussions}\label{sec:results}
\begin{figure*}[htb]
\center{
\includegraphics[width=0.9\textwidth]{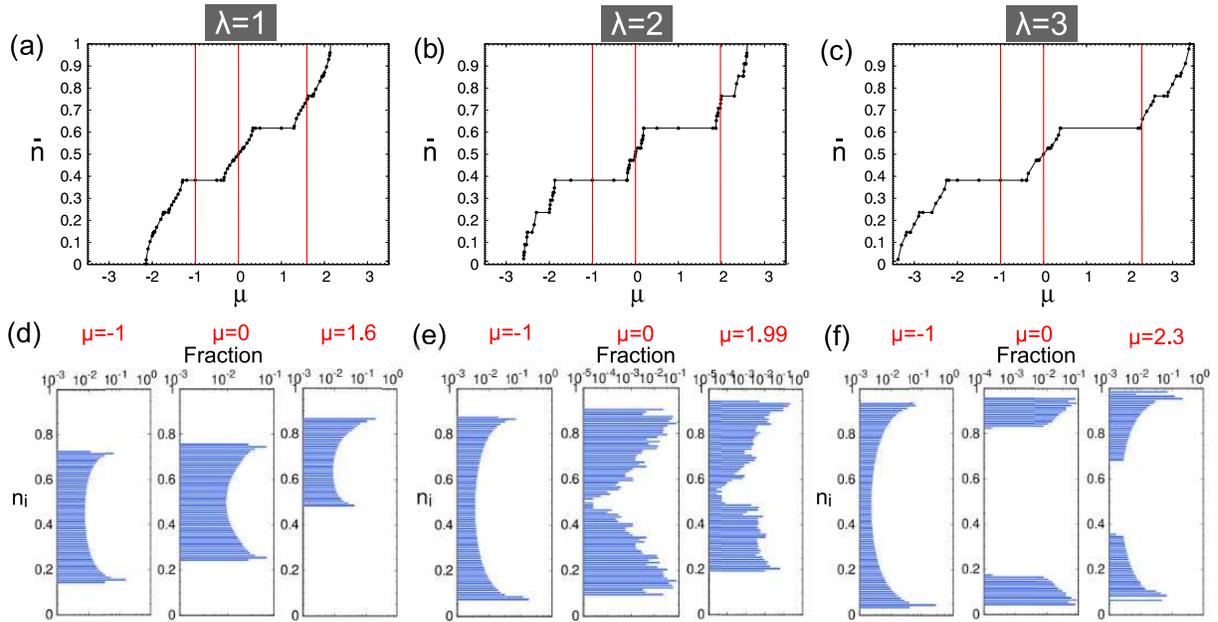}}
\caption{(a,b,c) Average density $\nb$ plotted against the chemical potential $\mu$ and (d,e,f) the histogram of $n_i$ for $\l=1$, 2, and 3, respectively.}
\label{fig:n-mu}
\end{figure*}

\subsection{Density of states and eigenfunctions}\label{ssec:dos}
We first review known results for the density of states (DOS) and the eigenfunctions of the AAH model,  showing calculated results.
As shown in Figs.~\ref{fig:aah}(a) and \ref{fig:aah}(d), when the quasiperiodic potential is weak ($\l<2$), the electron state is extended in real space and the DOS has a continuous spectrum (though it is separated by gaps).
At $\l=2$, eigenfunctions exhibit self-similar distributions and the DOS is singular continuous [Figs.~\ref{fig:aah}(b) and \ref{fig:aah}(e)].
For $\l>2$, eigenfunctions are localized and the DOS is a dense set of $\d$ functions. 
Note that the maximum (minimum) eigenvalue of the Hamiltonian (\ref{eq:aah}) at $\mu=0$ is $\Emax$ ($-\Emax$)  with $\Emax=2.1441$, 2.5975, and 3.3862 for $\lambda=1$, 2, and 3, respectively.
We hence vary $\mu$ only within $[-\Emax,\Emax]$ in the following.

\subsection{Charge distribution}\label{ssec:ni}
We find that the charge distribution $\{n_i\}$, which is a sum over eigenstates below the Fermi energy [i.e., Eq.~(\ref{eq:ni})], also changes its character with $\l$.
Figures \ref{fig:aah}(g-i) show the results for $\mu=0$.
At $\l=1$, $n_i$ continuously distributes from its minimum to the maximum [Fig.~\ref{fig:aah}(g)].
At $\l=2$, however, the population of $n_i$ decreases around the center of the distribution [Fig.~\ref{fig:aah}(h)].
At $\l=3$, the distribution bifurcates into roughly two values and shows a gap between them [Fig.~\ref{fig:aah}(i)].
Note that these (and following) results at $\l=1$ ($\l=3$) are representative of the results for $\l<2$ ($\l>2$) as we have obtained essentially the same results for various values of $\l<2$ ($\l>2$) though not explicitly shown.
 
\begin{figure}[tb]
\center{
\includegraphics[width=0.48\textwidth]{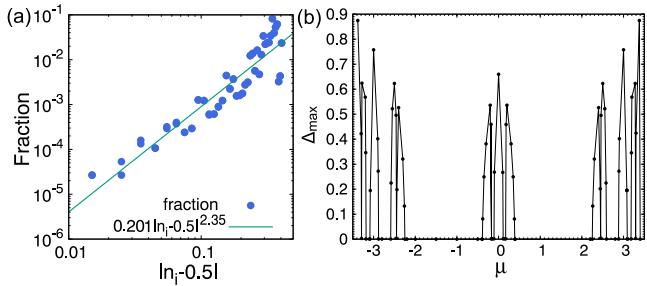}}
\caption{(a) Fraction of the $n_i$ plotted against $|n_i-n_c|$ for $\mu=0$ and $\l=2$ [corresponding to the middle histogram of Fig.~\ref{fig:n-mu}(e)]. The green line shows a linear fitting in the logarithmic scale. (b) Jump $\D_{\rm max}$ in the $n_i$ distribution for $\lambda=3$.}
\label{fig:jump}
\end{figure}

As $\mu$ increases, the average filling $\nb$ increases unless it is in the gap of the DOS, where $\nb$ does not change [Figs.~\ref{fig:n-mu}(a,b,c)].
Figures \ref{fig:n-mu}(d,e,f) show the histogram of $n_i$ at several values of $\mu$ denoted in Figs.~\ref{fig:n-mu}(a,b,c), respectively.
At $\l=1$, the distribution has
no jump in the histogram, irrespective of whether $\mu$ is located inside a spectral gap or not [Fig.~\ref{fig:n-mu}(d)].
When $\mu$ is located in a gap, the distribution 
has no jump even for $\l\geq 2$ [Figs.~\ref{fig:n-mu}(e)(f) left panel].
However, when $\mu$ is located at a support of the DOS, the distribution changes:
At $\l=2$, the population of $n_i$ becomes vanishingly small at a value, $n_c$ ($=0.5$ for $\mu=0$ for instance), and shows a power-law decay like $|n_i-n_c|^\gamma$ ($\gamma>0$) around it [Fig.~\ref{fig:n-mu}(e) middle and right panels].
In Fig.~\ref{fig:jump}(a), we fit the fraction for $\l=2$ and $\mu=0$ plotted against $ |n_i-0.5|$, to obtain $\g\simeq 2.35$.
 
At $\l=3$ the histogram shows a clear jump and bifurcates when $\mu$ is located at a support of the DOS [Fig.~\ref{fig:n-mu}(f) middle and right panels].
To quantify the jump, we sort $\{n_i\}$ for each $\mu$ in the ascending order and define the maximum difference between neighboring two values as $\D_{\rm max}$.
When this procedure is applied to the case of $\l\leq 2$,  $\D_{\rm max}$ is always negligibly small as expected from the continuous distributions in Figs.~\ref{fig:n-mu}(d) and \ref{fig:n-mu}(e): 
For $\l=2$, even when $\mu$ is located at a support of the DOS, the distribution continuously decreases and vanishes at a single point [Fig.~\ref{fig:n-mu}(e) middle and right panels], so that $\D_{\rm max}$ vanishes. 
However, for $\l=3$, when $\mu$ is located at a support of the DOS, the distribution has a jump [Fig.~\ref{fig:n-mu}(f) middle and right panels], so that $\D_{\rm max}$ is finite as plotted in 
Fig.~\ref{fig:jump}(b). 
Remarkably, the support of $\D_{\rm max}$ completely agrees with that of the DOS shown in Fig.~\ref{fig:aah}(c). 
Namely, we find a nontrivial relationship between the spectrum and the charge-distribution histogram. 

\subsection{Hyperuniformity}\label{ssec:hyper_results}

\begin{figure}[tb]
\center{
\includegraphics[width=0.48\textwidth]{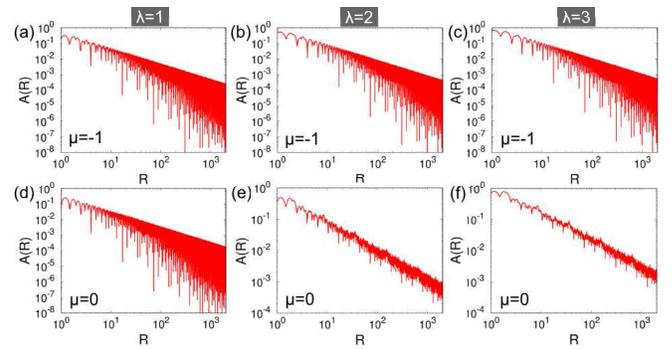}}
\caption{$A(R)$ calculated at $\mu=-1$ ($\mu=0$) for (a,b,c) [(d,e,f)] $\l=1$, 2, and 3, respectively. }
\label{fig:ar}
\end{figure}

We analyze these charge distributions in terms of hyperuniformity.
We first plot in Fig.~\ref{fig:ar} $A(R)$ of Eq.~(\ref{eq:ar}) for various values of $\l$ and $\mu$.
Irrespective of the potential strength $\l$ and whether $\mu$ is in a gap of the DOS, $A(R)$ always decreases in a power law and goes to zero in the large-$R$ limit.
Therefore, the charge distribution of the AAH model is always hyperuniform.
We point out here that this fact discriminates the AAH model from random systems, where the charge distribution is not hyperuniform (see Appendix \ref{sec:random}).

\begin{figure}[tb]
\center{
\includegraphics[width=0.48\textwidth]{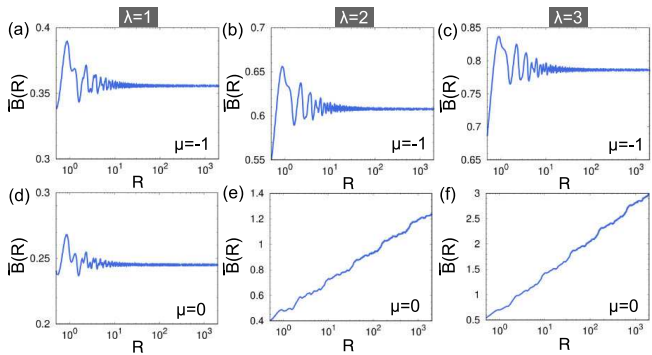}}
\caption{(a,b,c) [(d,e,f)] $\Bb(R)$ calculated at $\mu=-1$ ($\mu=0$) for $\l=1$, 2, and 3, respectively. }
\label{fig:br}
\end{figure}

\begin{figure}[tb]
\center{
\includegraphics[width=0.48\textwidth]{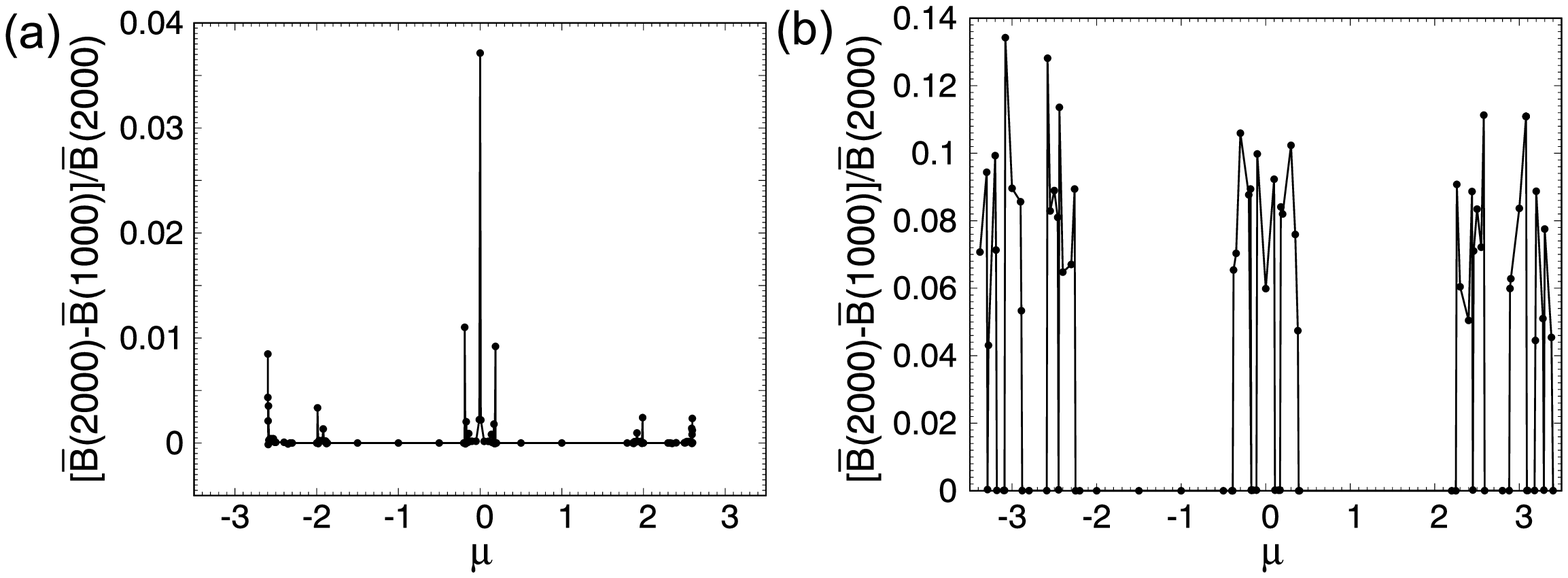}}
\caption{Difference between $\Bb$ evaluated at $R=1000$ and $R=2000$ (normalized by the latter value) for (a) $\l=2$ and (b) $\l=3$.}
\label{fig:omgrad}
\end{figure}

We then calculate $\Bb(R)$ of Eq.~(\ref{eq:br}) for the same parameters and plot them in Fig.~\ref{fig:br}.
In panels (a-d) $\Bb(R)$ converges to a constant value at a large $R$, which means that these distributions are Class-I hyperuniform \cite{torquato18}.
As we have seen in Fig.~\ref{fig:n-mu}, all these distributions have a 
histogram without a jump.

On the other hand, in Figs.~\ref{fig:br}(e) and \ref{fig:br}(f), $\Bb(R)$ increases logarithmically with $R$, which means that these distributions are Class-II hyperuniform \cite{torquato18}. 
We have confirmed this point with another calculation in momentum space, too (see Appendix \ref{sec:zk}).
As we have seen in Figs.~\ref{fig:n-mu} and \ref{fig:jump}, these distributions have a point or a region where the fraction in the histogram becomes zero:
For $\l=2$ and $\mu=0$, the fraction vanishes at $n_c=0.5$ while for $\l=3$ and $\mu=0$ the distribution has a finite jump in the histogram.
 
To quantify the above argument, we calculate the difference of $\Bb(R)$'s calculated at $R=1000$ and 2000.
For $\l<2$, this quantity is virtually zero while for $\l\geq 2$ it can be finite depending on $\mu$. 
Figures \ref{fig:omgrad}(a) and \ref{fig:omgrad}(b) show the results at $\l=2$ and 3, respectively.
For $\l=3$, the $\mu$ values giving a finite difference of $\Bb(R)$ completely agree with the $\mu$ values of a finite $\D_{\rm max}$ in Fig.~\ref{fig:jump}(b), as well as with the support of the DOS in Fig.~\ref{fig:aah}(c).
For $\l=2$, corresponding to the singular continuous spectrum in Fig.~\ref{fig:aah}(b), the difference of $\Bb(R)$ shows peaks of measure zero at the support of the DOS.
 
\begin{figure*}[tb]
\center{
\includegraphics[width=0.9\textwidth]{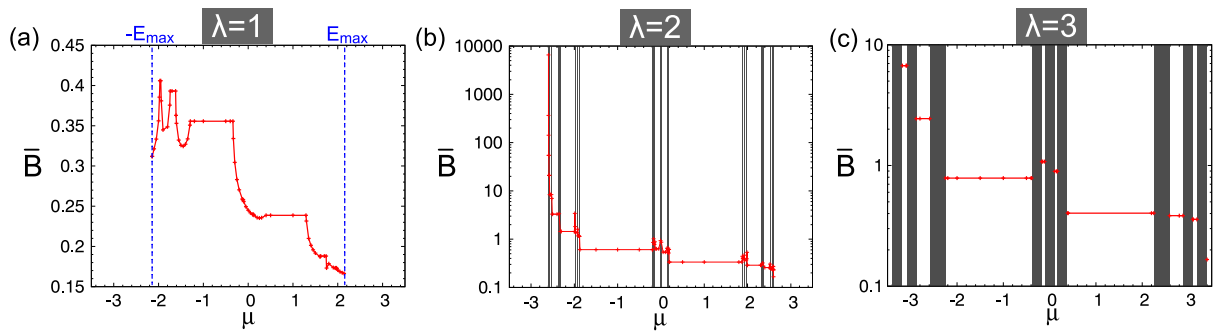}}
\caption{(a-c) $\Bb(R=2000)$ calculated for $\lambda=1$, 2, and 3, respectively, in the range $[-\Emax,\Emax]$. Black region represents the Class-II hyperuniformity, where $\Bb(R)$ is not well defined. In (a), $\pm\Emax$ are denoted by blue dashed lines.}
\label{fig:om-mu}
\end{figure*}

Thus, we obtain i) Class-I hyperuniformity for $\l<2$, ii) Class-I or II hyperuniformity for $\l\geq 2$, depending on the location of $\mu$ in the DOS.
For the Class-I hyperuniformity, the order metric $\Bb=\Bb(\infty)$ represents the degree of regularity.
Generally speaking, $\Bb$ is smaller for a simpler distribution \cite{torquato18}.

We evaluate $\Bb$ at $R=2000$ and plot it for $\l=1$, 2, and 3 in Fig.~\ref{fig:om-mu}.
For $\l=1$, $\Bb$ is always defined and relatively small. 
Since $n_i\equiv1$ for $\mu\geq\Emax$,  $\Bb$ at $\mu=\Emax$ agrees with that of the point distribution of the integer lattice, $1/6$  \cite{torquato18}.
On the other hand,  as $\mu$ approaches $-\Emax$, $\Bb$ goes to $\sim0.31$, which is consistent with the value obtained for the lowest-energy eigenstate in Ref.~\cite{sakai22}.
$\Bb$ changes significantly when $\mu$ moves within the support of the DOS, while it is constant for $\mu$ moving within a gap. 
As we see in Fig.~\ref{fig:aah}(a), the DOS has a sharp ($\d$-functional) peak at the gap edges.
A general trend is that the inclusion of the states around the upper edge of a gap reduces $\Bb$ significantly while the states around the lower edge of a gap increase $\Bb$ relatively less significantly.

For $\l=2$, the distribution is Class-II hyperuniform when $\mu$ is located on the support of the singular continuous DOS.
However, as $\mu$ crosses it, $\Bb$ changes significantly, while $\Bb$ is constant for $\mu$ inside a gap.
For $\mu=\Emax$,  $\Bb=1/6$ for the same reason as above.
On the other hand, $\Bb$ becomes extremely large for $\mu\to -\Emax$.
This is because the lowest-energy eigenstate at $\l=2$ is multifractal and is not hyperuniform.
Although other eigenstates are also multifractal, the charge distribution, which is a sum over many eigenstates below the Fermi level, is hyperuniform. 
As $\mu$ increases, more states contribute to $n_i$, making $\Bb$ tend to decrease.

For $\l=3$, the region of Class-II hyperuniformity expands, corresponding to the DOS in Fig.~\ref{fig:aah}(c).
In the Class-II region, the histogram of $n_i$ has a jump, as we have seen in Figs.~\ref{fig:n-mu}(f) and \ref{fig:jump}(b).
In other regions, the histogram has no jump
and $\Bb$ is well defined. It tends to decrease as $\mu$ increases across the support of the DOS, except for the region around $\mu=0$.
While $\Bb=1/6$ at $\mu=\Emax$, the region slightly above $\mu=-\Emax$ is Class-II hyperuniform. 
However, in the limit of $\mu\to-\Emax$, it is not hyperuniform \cite{sakai22} since the lowest-energy eigenstate is localized.

\begin{figure}[tb]
\center{
\includegraphics[width=0.48\textwidth]{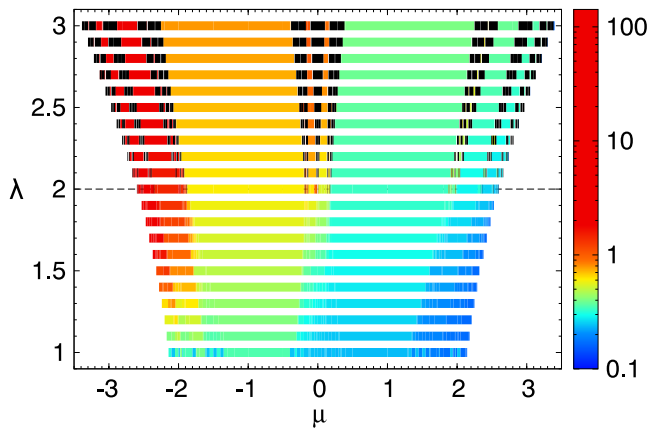}}
\caption{$\l-\mu$ diagram of the hyperuniformity classes and the order metric [$\Bb(2000)$] of the AAH model. Black region represents the Class-II hyperuniformity. The calculation was done for $-\Emax <\mu <\Emax$.}
\label{fig:diagram}
\end{figure}

Performing similar calculations for various values of $\mu$ and $\l$, we summarize the results of $\Bb$ and hyperuniformity class in Fig.~\ref{fig:diagram}.
The black region shows Class II while the color in other regions represents the order metric $\Bb$ of Class I.
A general trend is that $\Bb$ is larger for a smaller $\mu$ and larger $\l$.
In Appendix \ref{sec:fib}, we show that the Fibonacci models, where the eigenstates are always critical, show a behavior similar to the $\l=2$ case of the AAH model.

\subsection{Phase transitions and criticality}\label{ssec:phase}

\begin{figure*}[tb]
\center{
\includegraphics[width=0.95\textwidth]{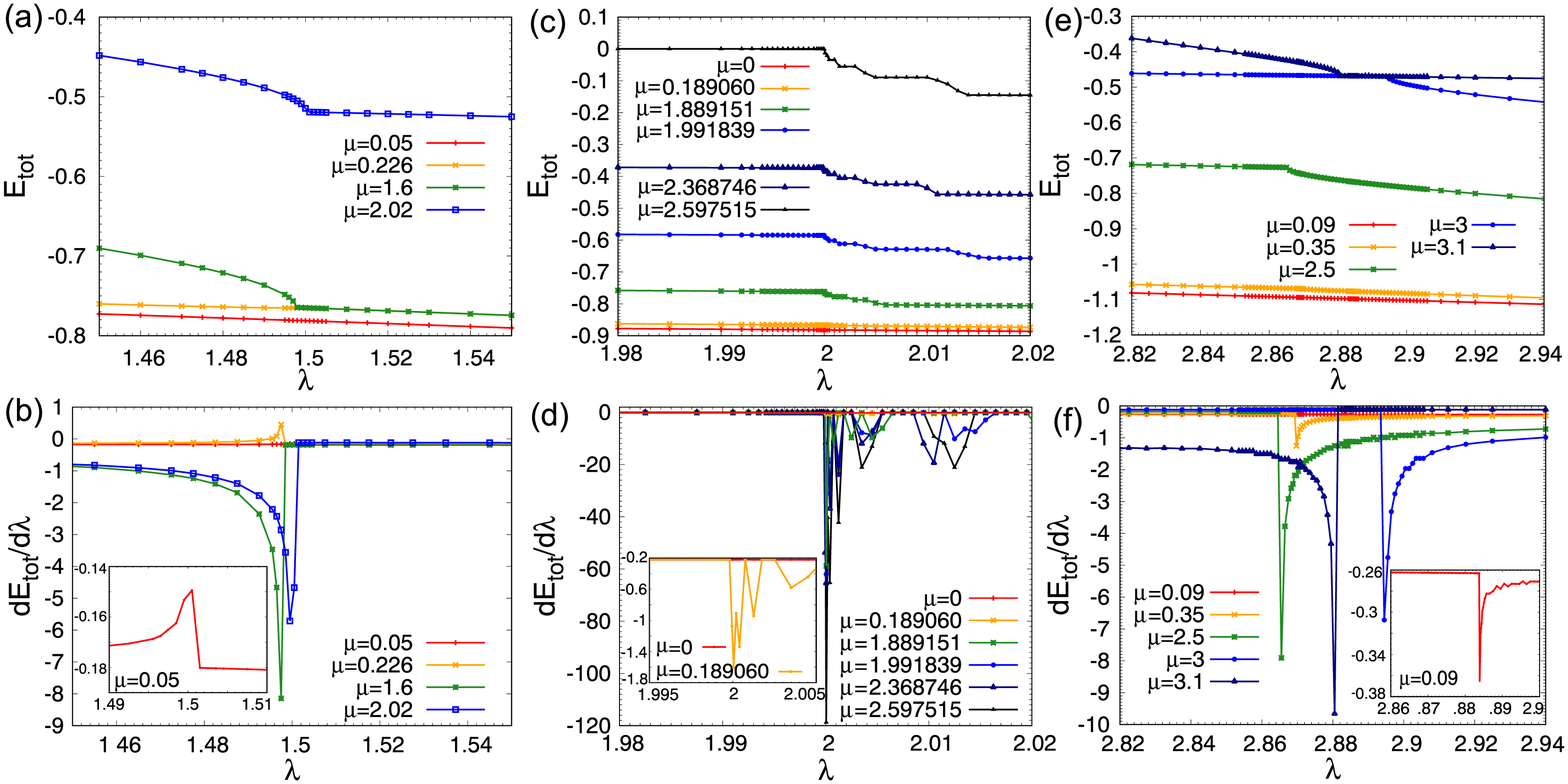}}
\caption{Total energy and its first derivative calculated around the phase boundaries; (a,b) $\l\sim1.5$, (c,d) $\l\sim2$, and (e,f) $\l\sim2.8$. Insets to (b,d,f) are enlarged views of the smallest $\mu$ data.}
\label{fig:etot}
\end{figure*}

\begin{figure}[tb]
\center{
\includegraphics[width=0.48\textwidth]{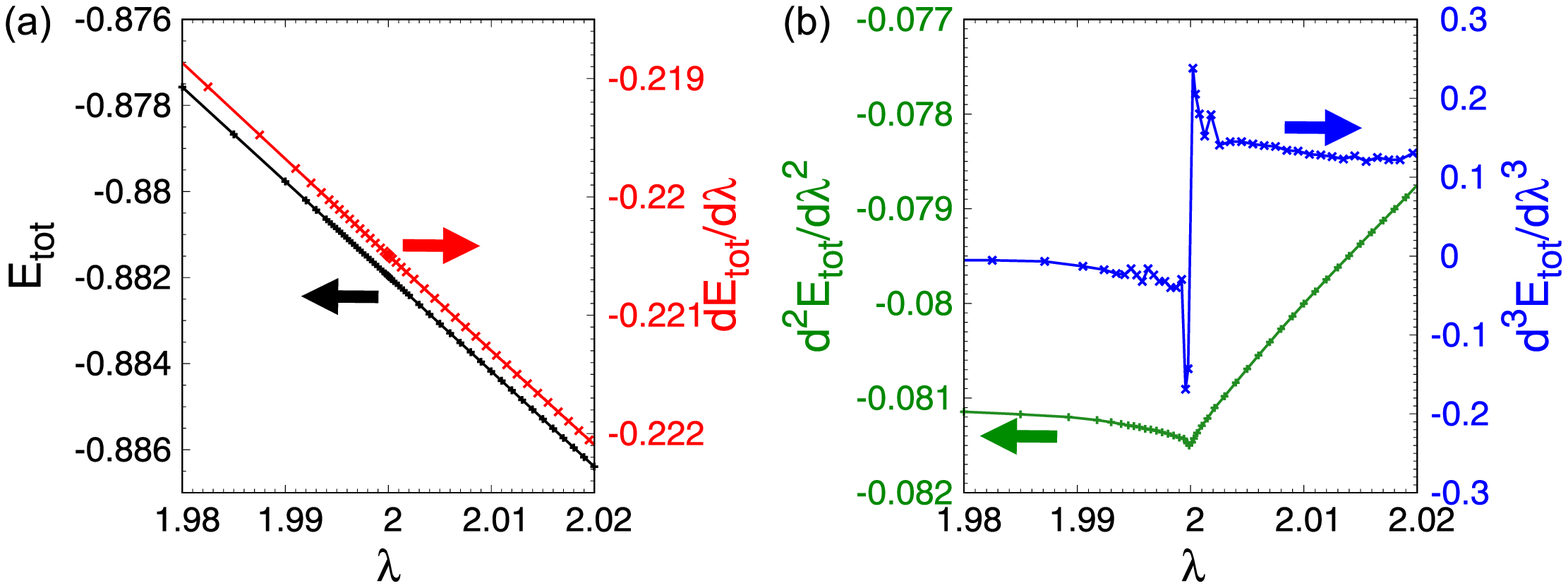}}
\caption{Total energy and its derivatives calculated at $\mu=0$ around $\l=2$.
(a) $\Etot$ and its first derivative. (b) The second and third derivatives.}
\label{fig:etot_mu0}
\end{figure}

As we have found in Fig.~\ref{fig:diagram}, the hyperuniformity class and order metric change with $\mu$ and $\l$. In particular, abrupt changes occur at the border of the Class-I and II regions. In this section, we examine whether these changes manifest themselves as a phase transition.
We numerically calculate the total energy,
\begin{align}
    \Etot\equiv\sum_{E_\a< 0} E_\a \label{eq:etot}
\end{align}
and its derivative (evaluated by a difference between neighboring two data points) with respect to $\l$ for fixed $\mu$'s at zero temperature.
Since the distribution of $\{E_\a\}$ is electron-hole symmetric, we concentrate only on the $\mu\geq0$ side.

First, for $\l<2$, the order metric in Fig.~\ref{fig:om-mu}(a) shows abrupt changes when $\mu$ crosses the gap edge.
Since the eigenstates are extended, this is a metal-insulator transition.
As a function of $\l$, too, $\Etot$ shows a kink and its first derivative shows a jump, as shown in Figs.~\ref{fig:etot}(a) and \ref{fig:etot}(b).
Here, we have chosen several $\mu$ values which show a singularity around $\l=1.5$.
The first derivative shows a rapid increase around the critical point presumably because of the large DOS at the gap edges in one dimension.
Aside from these singularities, $\Etot$ and $d\Etot/d\l$ curves are smooth, showing no phase transition, even though $\Bb$ changes.

For $\l>2$, on the other hand, all the eigenstates are localized, so that no metal-insulator transition occurs.
Nevertheless, $\Etot$ plotted against $\l$ still shows a kink, as shown in Fig.~\ref{fig:etot}(e), where we have chosen several $\mu$ values crossing the border of Class-I and -II regions in Fig.~\ref{fig:diagram}.
Notice that the relatively flat side corresponds to Class I.
The presence of the kink is evidenced in the plots of $d\Etot/d\l$ in Fig.~\ref{fig:etot}(f).
This transition may be viewed as a transition from a band insulator (in the sense that the DOS vanishes though a `band' is not well defined) to an Anderson insulator (where the DOS is finite but the mobility vanishes though the potential is not random but quasiperiodic).

Around $\l=2$, we need a more careful analysis because the DOS is singular continuous. We have fine-tuned the $\mu$ values to several eigenenergies at $\l=2$ and plotted $\Etot$ and $d\Etot/d\l$ in Figs.~\ref{fig:etot}(c) and \ref{fig:etot}(d), respectively.
We find kinks in $\Etot$ and jumps in $d\Etot/d\l$ at $\l\sim2$ for all the $\mu$ values except $\mu=0$. 
We see several additional kinks for $\l\gtrsim2$, which are due to the crossing of the eigenenergies with a very small measure in this region.

All the above results except for $\mu=0$ show the first-order transition between the gapped and ungapped regions.
On the other hand, at $\mu=0$, where the hyperuniformity class changes at $\l=2$, no jump is observed in $d\Etot/d\l$ [Fig.~\ref{fig:etot_mu0}(a)].
In fact, $\mu=0$ is special because the DOS does never vanish for any $\l$ due to the electron-hole symmetry and the self-duality.
We then calculate $d^2\Etot/d\l^2$ and $d^3\Etot/d\l^3$ (by a difference between neighboring two data points), plotting them in Fig.~\ref{fig:etot_mu0}(b).
We find that $d^2\Etot/d\l^2$ is still continuous but has a kink whereas $d^3\Etot/d\l^3$ shows a jump.
This weak singularity may be attributed to the singular continuous DOS at $\l=2$.
We have thus revealed a third-order criticality at $\l=2$ and $\mu=0$.

These results clarify whether and where a phase transition occurs between electronic states with different inhomogeneous but orderly charge patterns.
For $\l<2$, while the observed phase transition is  attributed to the metal-insulator one and is not so surprising, an important observation here is the {\it absence} of the phase transition in other regions where $\Bb$ (and hence the charge distribution) smoothly changes. 
For $\l> 2$, the phase transition occurs between two different {\it insulating} phases characterized by different hyperuniformity classes; no phase transition occurs within the same hyperuniformity class.
These results in turn prove an essential role of hyperuniformity analysis, which allows us to detect the phase transition in aperiodic systems independently of the total-energy calculation, like the role played by the order parameter in periodic systems. 

\subsection{Multihyperuniformity}\label{ssec:multi}
\begin{figure}[tb]
\center{
\includegraphics[width=0.48\textwidth]{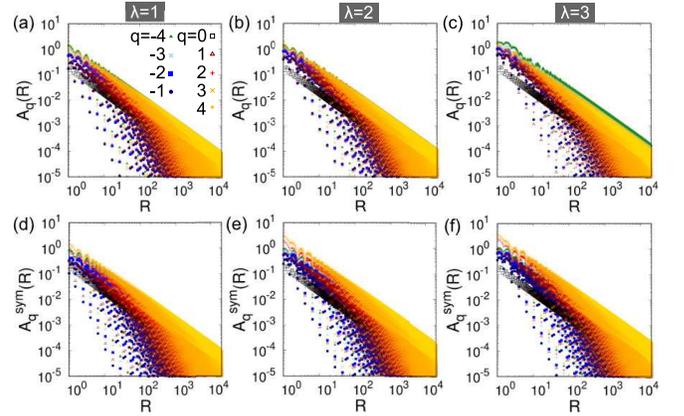}}
\caption{(a,b,c) $A_q(R)$ calculated for various $q$'s for $\mu=-1$ and $\l=1$, 2, and 3, respectively. (d,e,f) The same for $\Aqsym(R)$.}
\label{fig:aqr}
\end{figure}

\begin{figure}[tb]
\center{
\includegraphics[width=0.48\textwidth]{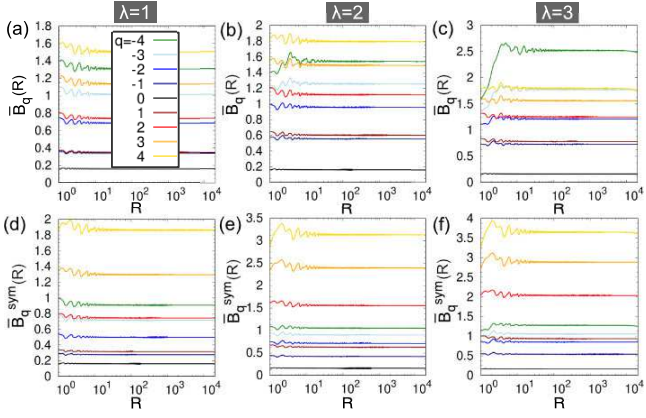}}
\caption{(a,b,c) $\Bbq$ plotted against $R$ for various $q$'s for $\mu=-1$ and $\l=1$, 2, and 3, respectively. (d,e,f) The same for $\Bbqsym$.}
\label{fig:bqr}
\end{figure}

\begin{figure*}[tb]
\center{
\includegraphics[width=0.9\textwidth]{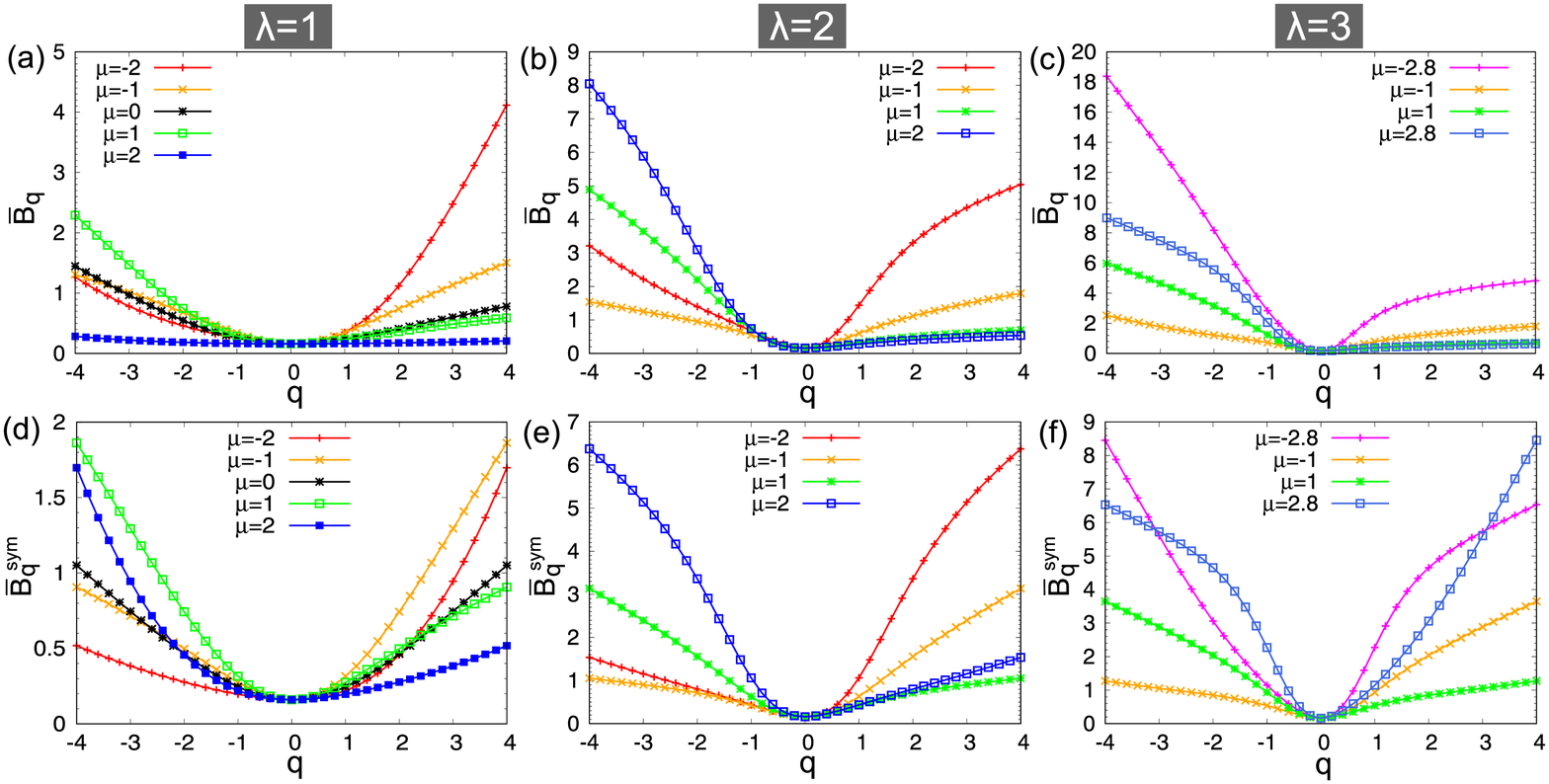}}
\caption{(a,b,c) $\Bbq$ calculated at $R=1000$ and $\mu=-1$ for $\l=1$, 2, and 3, respectively. (d,e,f) The same for $\Bbqsym$.}
\label{fig:bq}
\end{figure*}

\subsubsection{Straightforward extension}

So far the Class-I hyperuniform distributions have been characterized by just one scalar $\Bb$.
Here, with a simple extension of the definition (\ref{eq:nr}) of $N(R)$, we generalize the order metric to a function that should capture more detailed information on the density distribution. Namely, we define
\begin{align}
 N_q(R)\equiv\sum_{i=1}^{N} n_i^q \Theta(R-|r_i-r_c|), \label{eq:nqr}
\end{align}
and then $\s_q^2(R)$ in the same way as Eq.~(\ref{eq:s2}). 
In analogy with the multifractal dimension \cite{halsey86}, the exponent $q$ works as a filter to emphasize the contribution from a large (small) $n_i$ for $q>0 (<0)$.
Corresponding to Eqs.~(\ref{eq:ar}) and (\ref{eq:br}), we define
\begin{align}
    A_q(R) &\equiv  {\s_q}^2(R)/R,  \label{eq:aq}\\
   \Bbq(R) &\equiv\frac{1}{(\bar{n^{q}})^2 R} \int_0^R {\s_q}^2(R') dR'  \label{eq:bq}
\end{align}
with $\bar{n^q}\equiv \frac{1}{N}\sum_i n_i^q$.
By definition,  $\Bb_{q=0}(R)$ agrees with the order metric of the point distribution and $\Bb_{q=1}(R)$ agrees with $\Bb(R)$ of Eq.~(\ref{eq:br}).

While Eq.~(\ref{eq:nqr}) is a simple generalization, it would not be so obvious whether Eq.~(\ref{eq:nqr}) of $\{n_i^q\}$ gives a Class-I hyperuniformity (for which the order metric is well-defined) even when Eq.~(\ref{eq:nr}) of $\{n_i\}$ does.
We examine this point in Figs.~\ref{fig:aqr}(a-c) and \ref{fig:bqr}(a-c).
First, the former shows that $A_q(R)$ always goes to zero as $R$ increases, i.e., $\{n_i^q\}$ is also hyperuniform.
Then, the latter shows that $\Bbq(R)$ converges to finite values for all $q$'s in the large-$R$ limit.
Namely, $\{n_i^q\}$ is Class-I hyperuniform for all $q$'s.
We have obtained the same conclusion for other values of $\mu$ as far as $\{n_i\}$ belongs to Class I, as one may infer from the moderate values of $\Bbq$ in Fig.~\ref{fig:bq} below.
Note that when $\{n_i\}$ is Class-II hyperuniform, $\{n_i^q\}(q\neq 0)$ remains Class II for all the parameters we studied (Appendix \ref{sec:class2}).

In Figs.~\ref{fig:bq}(a-c), we plot $\Bbq$ (measured at $R=1000$) against $q$ for various $\l$ and $\mu$.
We find that $\Bbq$ takes the minimum at $q=0$, where $\Bbq$ agrees with the value ($1/6$) for a point distribution,
and is convex downward around $q=0$.
As $|q|$ increases, $\Bbq$ monotonically increases on each side of $q>0$ and $q<0$.
This reflects the larger spatial fluctuation for a larger $|q|$.

At $\l=1$, $\Bbq$ is larger on the $q<0$ ($q>0$) side for $\mu>0$ ($\mu<0$).
This is reasonable because for $\mu>0$ ($\mu<0$) small (large) values of $n_i^q$ can be further away from $\bar{n^q}$ (and hence more irregular) and $q<0$ ($q>0$) emphasizes these contributions. 
For $q>0$, $\Bbq$ tends to decrease as $\mu$ increases, as is expected from the behavior at $q=1$ displayed in Fig.~\ref{fig:om-mu}(a);
at $\mu=2$, all the sites are almost completely filled, so that $\Bbq$ is nearly flat for $q>0$.
For $q<0$, on the other hand,  $\Bbq$ shows a complicated dependence on $\mu$ though it should approach $1/6$ eventually for $\mu\to\Emax$.
In particular, the large $\Bbq$ for $\mu=1$ is interesting because this means that the charge distribution is significantly inhomogeneous even for this relatively large value of $\mu$.
In fact, as we see in Fig.~\ref{fig:n2}(a) in Appendix \ref{sec:n2}, the fluctuation measured by the local variance is maximized around $\mu=1$.

As $\l$ increases, $\Bbq$ tends to increase, reflecting the larger fluctuation and consequent irregularity, in particular on the $q<0$ side.
On the $q>0$ side, $\Bbq$ tends to decrease with $\mu$ in accord with Figs.~\ref{fig:om-mu}(b) and \ref{fig:om-mu}(c) for $q=1$.
$\Bbq$ shows a more complicated dependence on $\mu$ on the $q<0$ side.
It is interesting that $\Bbq$ at $\l=3$ is always larger for $q<0$ than for $q>0$.
For $\mu<0$,  this is opposite to what we have seen at $\l=1$.
This is presumably because $\{n_i\}$ for $\mu<0$ reflects more directly the structure of localized eigenfunctions, which have vanishingly small amplitudes at most sites.

\subsubsection{Symmetric definition}
In Fig.~\ref{fig:bq}(a), we see that $\Bbq$ at $\mu=0$ (black curve) is asymmetric with respect to $q=0$.
However, as the charge distribution at $\mu=0$ is symmetric with respect to $n_c=0.5$ [see Figs.~\ref{fig:aah}(g) and \ref{fig:n-mu}(d)], it may be preferable to define an order metric to reflect this symmetry.
The asymmetry of $\Bbq$ defined by Eq.~(\ref{eq:bq}) comes from the fact that $(0.5+\d)^q$ does not agree with $(0.5-\d)^{-q}$, where $\d$ represents a deviation from the average value 0.5.
Hence, to remedy this asymmetry, we define $s_i\equiv\sqrt{n_i/(1-n_i)}$ and
\begin{align}
 N_q^{\rm sym}(R)\equiv\sum_{i=1}^{N} {s_i}^q \Theta(R-|r_i-r_c|). \label{eq:nqrsym}
\end{align}
Notice that $s_i$ at $n_i=0.5+\d$ equals $s_i^{-1}$ at $n_i=0.5-\d$. 
Then, we define ${\s_q^{\rm sym}}^2(R)$ in the same way as Eq.~(\ref{eq:s2}) and
\begin{align}
    \Aqsym(R) &\equiv  {\s_q^{\rm sym}}^2(R)/R,  \label{eq:aqsym}\\
   \Bbqsym(R) &\equiv\frac{1}{(\bar{s^q})^2 R} \int_0^R {\s_q^{\rm sym}}^2(R') dR'  \label{eq:bqsym}
\end{align} 
with $\bar{s^q}\equiv\frac{1}{N}\sum_i s_i^q$ . 
$\Bb_{q=0}^{\rm sym}(R)$ agrees with the order metric of the point distribution and $\Bbqsym(R)$ is symmetric with respect to the transformation $(\mu,q)\leftrightarrow(-\mu,-q)$ as far as the DOS for $\mu=0$ is symmetric with respect to $\w=0$.

As was done above, we first check the large-$R$ behavior of $\Aqsym(R)$ in Figs.~\ref{fig:aqr}(d-f).
The results show that $\{s_i^q\}$ is hyperuniform for all the $q$ values studied.
We then plot in Figs.~\ref{fig:bqr}(d-f) the corresponding $\Bbqsym(R)$ against $R$. 
We find that $\{s_i^q\}$ belongs to Class I for all the parameters for which $\{n_i\}$ belongs to Class I.
We have obtained the same conclusion for all other choices of $\mu$ that we study though not shown.
Note that, when $\{n_i\}$ belongs to Class II, $\{s_i^q\}(q\neq 0)$ also shows Class-II behavior (Appendix  \ref{sec:class2}).

We plot $\Bbqsym$ measured at $R=1000$ in Figs.~\ref{fig:bq}(d-f).
First, for $\l=1$ and $\mu=0$ (black curve), we see that the curve is symmetric with respect to $q \leftrightarrow -q$, as expected.
$\Bbqsym$ takes the minimum of 1/6 at $q=0$.
Second, all the curves are symmetric against the simultaneous sign reversal of $\mu$ and $q$, i.e., $(\mu,q)\leftrightarrow(-\mu,-q)$.  
Therefore, the asymmetry of the $\Bbqsym$ curves for $\mu\neq 0$ correctly represents the asymmetric distribution of $\{n_i\}$ around $\nb$.

Another notable difference from the $\Bbq$ curves is that the $\Bbqsym$ curves do not approach a flat curve for $\mu\to\Emax$
(see blue curves).
This is due to the denominator of $\sqrt{n_i/(1-n_i)}$, which amplifies more the sites closer to $n_i=1$.
Namely, $\Bbqsym$ for $\mu\to\Emax$ reflects the structure of the highest-energy eigenfunction, just as $\Bbq$ for $\mu\to-\Emax$ does for the lowest-energy eigenfunction.
Note that $\Bbqsym$ for $\mu\to-\Emax$ still reflects the structure of the lowest-energy eigenfunction though its contribution to $\Bbqsym$ differs from that to $\Bbq$ due to the difference between $n_i$ [in Eq.~(\ref{eq:nqr})] and $s_i\sim\sqrt{n_i}$ [in Eq.~(\ref{eq:nqrsym})] in this region.

At $\l=1$, $\Bbqsym$ is larger for $q<0$ ($q>0$) for $\mu>0$ ($\mu<0$) for the same reason described above for $\Bbq$.
The same occurs for $\l=2$ and even for $\l=3$ and $\mu=\pm 1$.
For $\l=3$ and $\mu=\pm2.8$, while the same occurs for $|q|\lesssim3$, it is reversed for $|q|\gtrsim3$.
This is likely because the structure of the highest- or lowest-energy eigenstates (rather than the filling controlled by $\mu$) becomes more relevant for $\mu$ close to $\pm \Emax$ as mentioned above.

In Appendix \ref{sec:fib}, we calculate $\Bbqsym$ for the Fibonacci models. The convex-down behavior around $q=0$ and a monotonic increase with $|q|$, as well as a large enhancement at $\mu$'s close to $\pm\Emax$, are common to the Fibonacci models.

\subsubsection{Application to the critical regions}
\begin{figure}[tb]
\center{
\includegraphics[width=0.48\textwidth]{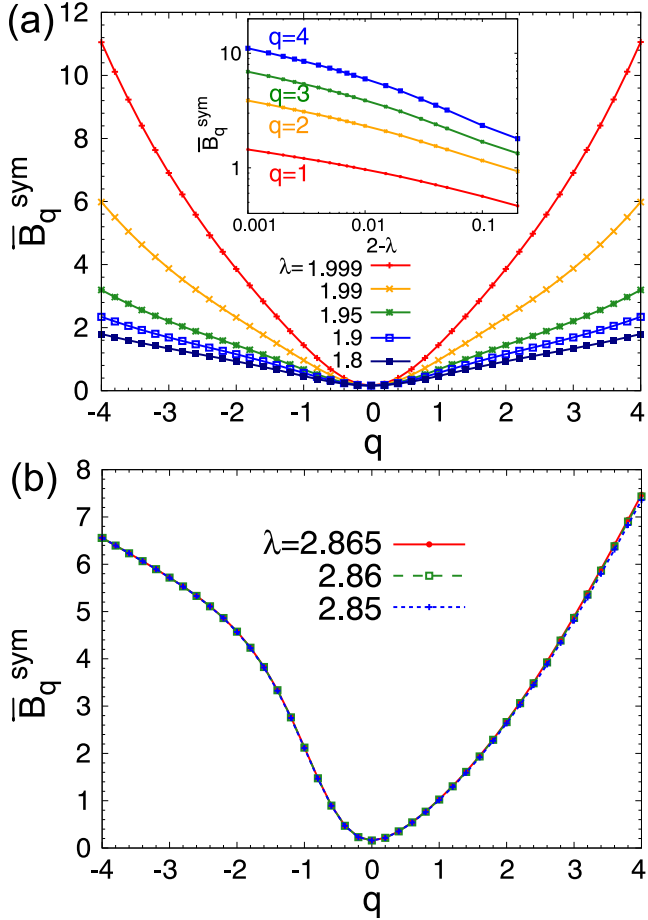}}
\caption{$\Bbqsym$ calculated around the phase transitions. (a) Around the third-order phase transition at $\l=2$ and $\mu=0$. Inset shows a plot against $2-\l$ in a logarithmic scale.
(b) Around the first-order phase transition at $\l\simeq2.866$ and $\mu=2.5$. The three curves are almost completely overlapping. Note that the distribution is Class-II hyperuniform for $\l\geq2.866$.}
\label{fig:bq_mu0}
\end{figure}

Here, we apply the multihyperuniformity analysis to a critical behavior around the phase transition discussed in Sec.~\ref{ssec:phase}. Our aim is to clarify how the inhomogeneous charge distribution changes around the critical point, by quantifying it through the generalized order metric.

In Fig.~\ref{fig:bq_mu0}(a), we focus on the continuous transition point at $\l=2$ and $\mu=0$.
Since the order metric is defined only in the Class-I hyperuniform region, we calculate $\Bbqsym$ only for $\l< 2$.
We find a rapid increase of $\Bbqsym$ for large $|q|$'s as $\l$ approaches the critical point.
This behavior means an increasing irregularity of the sites with a particularly large or small electron density.
Notice that $\Bb$ [Eq.~(\ref{eq:br})] alone cannot distinguish such a behavior from an overall increase of irregularity. 
In the inset, we plot $\Bbqsym$ against $2-\l$ in a logarithmic scale for several values of $q$. For each $q$, $\Bbqsym$ increases in a power law as $\l$ approaches the critical point, 2. The power seems to weakly depend on $q$, e.g., $-0.166$ at $q=1$ and $-0.248$ at $q=4$ for $2-\l<0.005$.

By contrast, Fig.~\ref{fig:bq_mu0}(b) shows that $\Bbqsym$ does not change on the Class-I side of the first-order phase transition at $\l\simeq2.866$ and $\mu=2.5$. The three curves are almost completely overlapping here. This of course means no significant change in the charge distribution up to the transition point and a jump there.

Our generalization thus offers a useful tool to analyze inhomogeneous density distributions, which are not multifractal but hyperuniform, and their changes by quantifying the irregularity of each contribution.

\section{summary and perspectives}\label{sec:summary}

\begin{table}
\centering{
\caption{Summary of the results obtained in Secs.~\ref{ssec:dos}, \ref{ssec:ni}, and \ref{ssec:hyper_results}.}\label{table:summary}
\begin{tabular}{c|c|c|c|c}
                                   & DOS at $\w=0$  & Distribution of $n_i$ & Hyperuniformity class \\ \hline
 \multirow{2}{*}{$\l<2$} &  0          &  No jump              & I\\
                                   &  $\neq0$      &  No jump             & I \\ \hline
  \multirow{2}{*}{$\l=2$} & 0          &  No jump            & I\\
                                    &   $\neq0$    & $|n_i - n_c|^\g (\g>0)$  & II\\ \hline
  \multirow{2}{*}{$\l>2$}  &  0         & No jump              & I\\
                                 &   $\neq0$        & Bifurcated by a jump  & II\\ \hline  
\end{tabular}}
\end{table}

We have studied the charge distribution in the Aubry-Andr\'e-Harper model in light of hyperuniformity.
According to the strength $\l$ of the quasiperiodic potential, the model is known to exhibit extended, critical, and localized electron states.
In this paper, we have revealed that the inhomogeneous distribution of electron charge $n_i$, which is neither periodic nor multifractal but still orderly, also changes its character with $\l$.
The character is quantified in the framework of hyperuniformity generalized to density distributions.

First, we have found a nontrivial relationship between $\l$, the DOS at the Fermi level, $\{n_i\}$, and the hyperuniformity class, as summarized in Table \ref{table:summary}.
For $\l<2$, where eigenstates are extended, the charge distribution 
has no jump and is Class-I hyperuniform.
There is no phase transition as far as the order metric changes smoothly while a first-order metal-insulator transition occurs in concomitance with an abrupt change of the order metric when the Fermi level $\mu$ crosses the gap edge.
For $\l> 2$, where eigenstates are localized, the charge distribution 
has no jump and Class-I hyperuniform only when $\mu$ resides in the gap of the DOS; otherwise, the charge distribution has a jump in its histogram and belongs to Class II.
While all the electron states are insulating in this region, the change from Class I to II manifests itself as a first-order phase transition.
At $\l=2$, where eigenstates are critical, the charge distribution 
has no jump and Class-I hyperuniform only when $\mu$ is in the gap of the DOS; otherwise, it shows a behavior vanishing at a single point in the histogram and belongs to Class II.
The transition is of the third order at $\mu=0$ and the first order otherwise.
For the Class-I hyperuniform distributions, we have also revealed the dependence of the order metric on $\l$ and $\mu$.

The hyperuniform charge distributions for $\l>2$ discriminate the AAH model from random systems, where the eigenstates are localized but the charge distribution is not hyperuniform (Appendix \ref{sec:random}). 
In addition to this, the eigenstates for $\l<2$ 
are also hyperuniform in the AAH model \cite{sakai22}.
These facts may make a significant difference between the localization-delocalization transition at $\l=2$ in the AAH model and the Anderson transition discussed in random systems in higher dimensions.

Since various extensions  \cite{sokoloff80,boers07,biddle10,biddle11,ganeshan15,sun15,gopalakrishnan17,devakul17,sutradhar19,szabo20} have been proposed for the AAH model, it is intriguing to explore these models in light of hyperuniformity. Of particular interest is the coexistence of localized and extended states at the same quasiperiodic potential observed in several models preserving a self-duality. The hyperuniformity analysis of the charge distribution in these models constitutes an important future issue.


Although the order metric seems to represent well a regularity of the aperiodic density distributions, it is obvious that much information about the distribution is lost in this quantification.
We therefore extend the order metric to a function, in analogy with the extension of the fractal dimension to the multifractal one \cite{halsey86}.
In both the straightforward extension and a symmetric definition, we first confirm that the order-metric function is well defined, i.e., $\{n_i^q\}$ and $\{s_i^q\}$ belong to Class I when $\{n_i\}$ belongs to Class I.
Thanks to the filtering effect of the power $q$, the order-metric function, $\Bbq$ or $\Bbqsym$, represents the regularity of differently weighted subsets of $\{n_i\}$. In particular, $\Bbqsym$ can correctly capture the asymmetry of the distribution.

This generalization applies to any density distribution ranging from 0 to 1 (i.e., probability distribution). 
As mentioned in the introduction, there are various density distributions, which are known to be neither random nor multifractal, on quasicrystalline structures.
Some of them may be hyperuniform. 
For instance, when an electron property on a quasiperiodic lattice is determined by short-range physics, it is likely hyperuniform.
To analyze such distributions, the generalized order-metric function will be a useful tool. 

\begin{acknowledgments}
This work was supported by JSPS KAKENHI Grant No. JP22H04603, JP19H00658, JP19H05825, and JP22H05114.
\end{acknowledgments}

\appendix 

\section{Comparison with Fibonacci models}\label{sec:fib}
\begin{figure}[tb]
\center{
\includegraphics[width=0.48\textwidth]{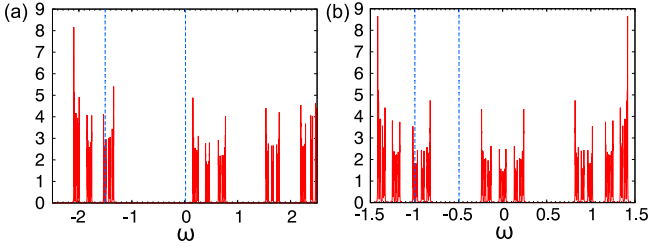}}
\caption{Density of states of the Fibonacci model. (a) Diagonal model with $V=t=1$ and $\mu=0$. (b) Off-diagonal model with $t_S=2t_L=1$ and $\mu=0$. Blue dashed lines indicate the chemical potentials used in Fig.~\ref{fig:br_fib} below.}
\label{fig:dos_fib}
\end{figure}

\begin{figure}[tb]
\center{
\includegraphics[width=0.48\textwidth]{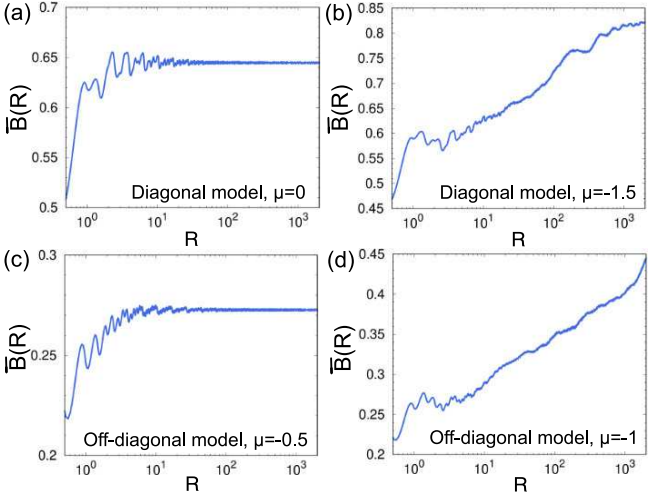}}
\caption{(a,b) $\Bb$ plotted against $R$ for the diagonal Fibonacci model with $V=t=1$ and $\mu=0$ and $-1.5$, respectively.
(c,d) The same for the offdiagonal Fibonacci model with $t_S=2t_L=1$ and $\mu=-0.5$ and $-1$, respectively.}
\label{fig:br_fib}
\end{figure}

\begin{figure}[tb]
\center{
\includegraphics[width=0.48\textwidth]{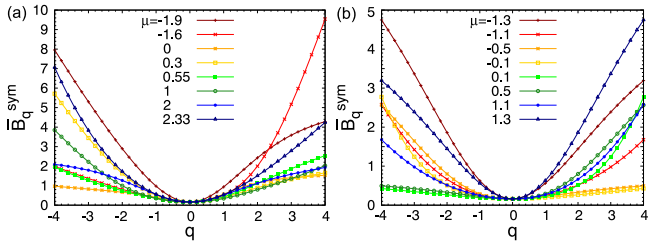}}
\caption{$\Bbqsym$ calculated for (a) the diagonal Fibonacci model with $V=t=1$, and (b) the off-diagonal Fibonacci model with $t_S=2t_L=1$. The values of $\mu$ are chosen to be inside eight major  gaps in the DOS.}
\label{fig:bqsym_fib}
\end{figure}

The Fibonacci models are known to exhibit critical eigenstates for any finite strength of quasiperiodic modulations \cite{kohmoto87,sutherland87,tokihiro88,mace17,jagannathan20}.
This behavior of the eigenstates corresponds to $\l=2$ in the AAH model.
One may therefore expect that the charge distribution in the Fibonacci models is Class-I hyperuniform when the chemical potential resides in a gap of the DOS, and Class-II hyperuniform otherwise. 

We examine the above expectation for the following two types of the Fibonacci model.\\
{\it Diagonal model:}
\begin{align}
 H_{\rm diag.}=-t\sum_i \left(\cc_{i+1}^\dagger \cc_i+{\rm h.c.}\right)+\sum_i \left(V_i-\mu\right)\cc_{i}^\dagger \cc_i,\label{eq:diag}
\end{align}
where $V_i=+V$ or $-V$ according to the Fibonacci sequence.\\
{\it Off-diagonal model:}
\begin{align}
 H_{\rm offdiag.}=-\sum_i t_i \left(\cc_{i+1}^\dagger \cc_i+{\rm h.c.}\right)-\mu\sum_i \cc_{i}^\dagger \cc_i,\label{eq:offdiag}
\end{align}
where $t_i=t_L$ or $t_S$ according to the Fibonacci sequence.\\
We numerically diagonalize the Hamiltonian for $N=F_{24}=75025$ sites under periodic boundary conditions.

For $\mu=0$, these models show the DOS of Figs.~\ref{fig:dos_fib}(a) and \ref{fig:dos_fib}(b), respectively.
We see that the DOS at the Fermi level ($\w=0$) is zero for $\mu=0$ in the diagonal model with $V=t=1$ and for $\mu=-0.5$ in the off-diagonal model with $t_S=2t_L=1$.
On the other hand, $\mu=-1.5$ in the diagonal model and $\mu=-1$ in the off-diagonal model are very close to the support of the DOS, whose measure is zero.

After confirming that $A(R)$ of Eq.~(\ref{eq:ar}) goes to zero in the large-$R$ limit, we plot in Fig.~\ref{fig:br_fib} $\Bb$ of Eq.~(\ref{eq:br}) against $R$. We find Class-I hyperuniformity for $\mu=0$ in the diagonal model [panel (a)] and $\mu=-0.5$ in the off-diagonal model [panel (c)].
The other two cases [panels (b) and (d)] show Class-II hyperuniformity. Note that a possible deviation from the expected $\log R$ behavior at large $R$ is attributed to the slight deviation of $\mu$ from the support of the DOS.
These results are fully consistent with those obtained for the AAH model at $\l=2$.

A recent study \cite{rai21} of the Fibonacci model revealed that the charge-density oscillation in the perpendicular space is related to the topological property when $\mu$ resides in a gap. Its relation with the Class-I hyperuniformity in the physical space is an interesting subject of future research.

In Fig.~\ref{fig:bqsym_fib}, we plot $\Bbqsym$ of Eq.~(\ref{eq:bqsym}) for the (a) diagonal and (b) off-diagonal models, where we select $\mu$ values residing eight major gaps in the DOS of Fig.~\ref{fig:dos_fib}.
All the curves take the minimum at $q=0$ and are convex downward around it, similarly to the results for the AAH model [Figs.~\ref{fig:bq}(d-f)]. 
We also see that $\Bbqsym$ tends to be large for $\mu$ close to $\pm\Emax$.
For the diagonal model, $\Bbqsym$ shows a complicated dependence on $\mu$. This would be at least partly due to the asymmetry in the DOS. 

For the off-diagonal model, on the other hand, $\Bbqsym$ shows a symmetry with respect to the exchange of $(\mu,q)$ and $(-\mu,-q)$, due to the electron-hole symmetry of the DOS. 
Interestingly, for $\mu>0 (<0)$, $\Bbqsym$ is larger on the $q>0(<0)$ side, on the contrary to the behavior in the AAH model at $\l=2$ [Fig.~\ref{fig:bq}(e)], suggesting a large irregularity of the eigenstates.
In addition, the $\mu=0.1$ and 0.5 curves show a similar behavior to each other. This may be related to a self-similarity of the model, because the gaps around these two $\mu$ points are related by a self-similar transformation.

\section{Comparison with a random system}\label{sec:random}
\begin{figure}[tb]
\center{
\includegraphics[width=0.48\textwidth]{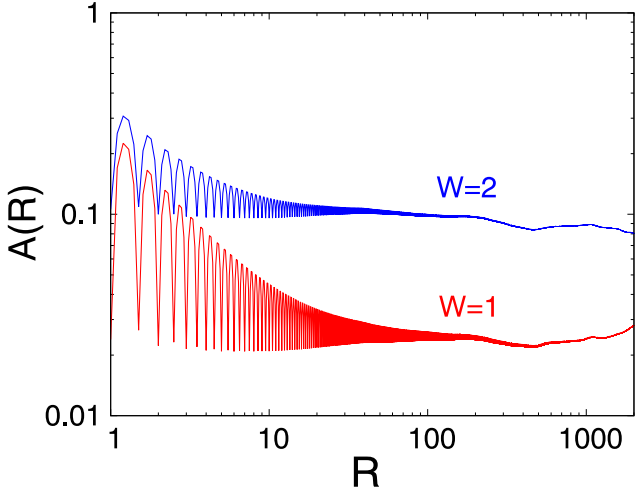}}
\caption{$A(R)$ of Eq.~(\ref{eq:ar}) calculated for the Hamiltonian (\ref{eq:random}) with  $W=1$ and 2.}
\label{fig:ar_random}
\end{figure}

Here, we demonstrate that the charge distribution in the localized phase in a random system is not hyperuniform.
We consider the following one-dimensional Anderson model\cite{anderson58},
\begin{align}
 H_{\rm random}=-t\sum_i \left(\cc_{i+1}^\dagger \cc_i+{\rm h.c.}\right)
   +\sum_i (W_i-\mu) \cc_{i}^\dagger\cc_i,\label{eq:random}
\end{align}
where $W_i$ is a random potential independently and uniformly distributed in the range $[-\frac{W}{2},\frac{W}{2}]\, (W>0)$.
All the states are localized for $W\ne 0$ \cite{abrahams79,anderson80}.
We numerically diagonalize the above Hamiltonian for 50000 sites and calculate the charge density at each site based on Eq.~(\ref{eq:ni}).
We then calculate $A(R)$ of Eq.~(\ref{eq:ar}) for the charge distribution.

The results for $W=1$ and 2 are plotted in Fig.~\ref{fig:ar_random}. 
We see that $A(R)$ remains finite at a large $R$.
This means that the charge distribution of the model (\ref{eq:random}) is not hyperuniform, unlike that of the AAH model.

The above results show that even in the localized ($\l>2$) region of the AAH model, there is a significant difference from the random system in light of the hyperuniformity of the charge distribution:
In the AAH model, it is either Class-I or II hyperuniform while it is not hyperuniform in a random system.
This difference may be used to distinguish a localization in quasiperiodic systems from that in random systems experimentally.

\section{Integrated intensity function}\label{sec:zk}

\begin{figure}[tb]
\center{
\includegraphics[width=0.48\textwidth]{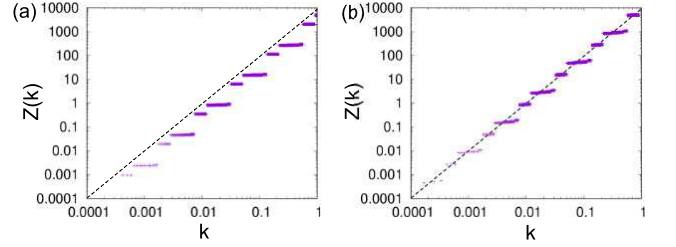}}
\caption{$Z_k$ calculated for (a) $\lambda=2$ and (b) 3. Black dashed lines correspond to the scaling of $\a=1$.}
\label{fig:zk}
\end{figure}

Here, we study the behavior of the structure factor,
\begin{align}
  S(k)=\left|\frac{1}{N}\sum_j n_j e^{-i k j}\right|^2-\nb^2\d(k),
\end{align}
at the long-wavelength limit ($k\to 0$).
The asymptotic behavior, $S(k)\sim k^\a$ for $k\sim 0$, is characterized by $\a>1$ for a Class-I and $\a=1$ for a Class-II hyperuniformity \cite{oguz17}.
Because this classification based on $\a$ does not rely on a window used in Sec.~\ref{ssec:hyper}, it gives an independent check for the determination of the hyperuniformity classes.
For quasiperiodic systems, where $S(k)$ consists of a dense set of Bragg peaks, an integrated intensity function,
\begin{align}
  Z(k)=2\int_0^k S(k) dk,
\end{align}
is smoother and hence more useful than $S(k)$ \cite{oguz17}.
Because $Z(k)$ behaves as $k^{\a+1}$ for $k\sim 0$, we plot it for (a) $\l=2$ and $\mu=0$ and (b) $\l=3$ and $\mu=0$ in a logarithmic scale in Fig.~\ref{fig:zk}.
We see that the results are consistent with $\a=1$ in both cases, supporting that the charge distributions for these parameters are Class-II hyperuniform.

\section{$\Bbq$ and $\Bbqsym$ for Class-II hyperuniform distributions}\label{sec:class2}
\begin{figure}[htb]
\center{
\includegraphics[width=0.48\textwidth]{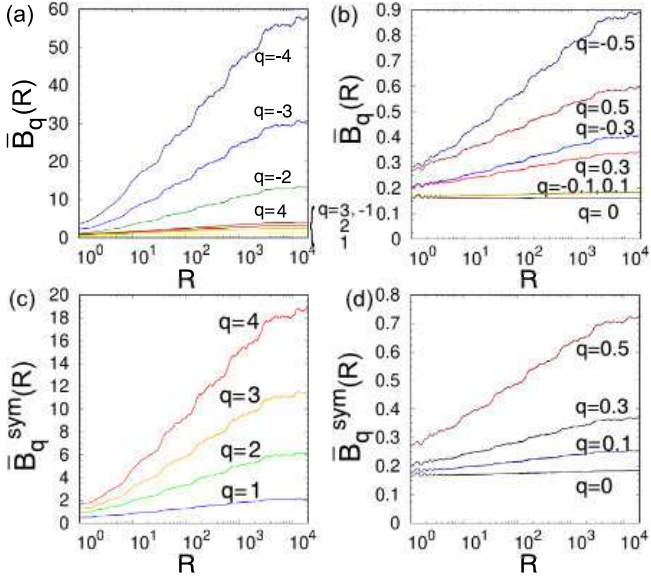}}
\caption{(a) $\Bbq$ plotted against $R$ for various $q$'s at the critical point ($\l=2$), where ${n_i}$ is Class-II hyperuniform. (b) The same plot for smaller $q$'s. (c), (d) The same as (a), (b) but for $\Bbqsym$.}
\label{fig:bqr2}
\end{figure}

\begin{figure}[htb]
\center{
\includegraphics[width=0.48\textwidth]{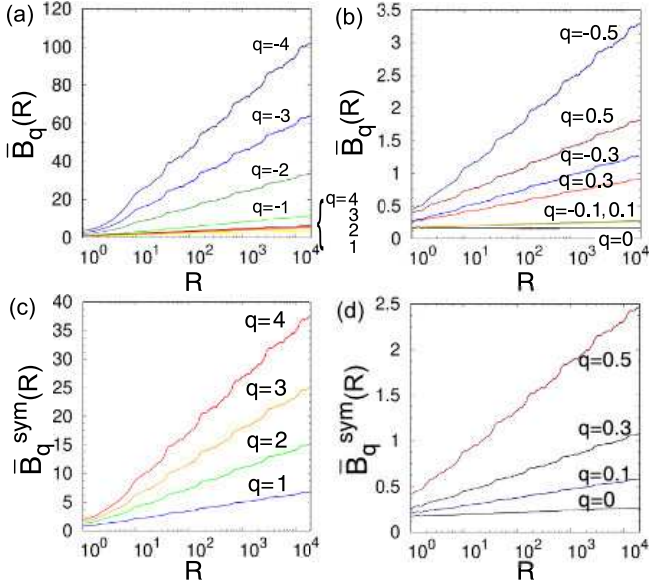}}
\caption{The same as Fig.~\ref{fig:bqr2} but for $\l=3$.}
\label{fig:bqr3}
\end{figure}

In Sec.\ref{ssec:multi}, we have shown that, when $\{n_i\}$ is Class-I hyperuniform, $\{n_i^q\}$ and $\{s_i^q\}$ also belong to Class I.
Here, we examine whether $\{n_i^q\}$ and $\{s_i^q\}$ are Class-II hyperuniform when $\{n_i\}$ is Class II.
After confirming that $A_q(R)$ and $\Aqsym(R)$ go to zero for $R\to\infty$, we plot $\Bbq(R)$ and $\Bbqsym(R)$ in Figs.~\ref{fig:bqr2} (for $\l=2$) and \ref{fig:bqr3} (for $\l=3$).
In both cases, we see that both $\Bbq(R)$ and $\Bbqsym(R)$ show Class-II behavior for $q\neq 0$.
Here, $\Bbqsym(R)$ is plotted only for $q\geq 0$ because of the symmetry.
Note that for $q=0$, both $\{n_i^q\}$ and $\{s_i^q\}$ are Class-I hyperuniform, where $\lim_{R\to\infty}\Bbq(R)$ and $\lim_{R\to\infty}\Bbqsym(R)$ agree with the order metric of the point distribution (i.e., $1/6$). 
As $|q|$ decreases, the gradient in the semi-logarithmic plots decreases while it seems that a finite positive gradient remains even for $|q|=0.1$.

\section{Local variance}\label{sec:n2}

\begin{figure}[tb]
\center{
\includegraphics[width=0.48\textwidth]{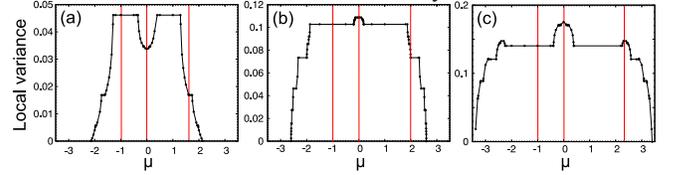}}
\caption{(a,b,c) The local variance for $\l=1$, 2, and 3, respectively. The red lines denote the values of $\mu$ presented in Fig.~\ref{fig:n-mu}.}
\label{fig:n2}
\end{figure}

One possible way to quantify the inhomogeneous charge distribution is to calculate the local variance defined by $\frac{1}{N}\sum_i(n_i-\nb)^2$.
This quantifies a local density fluctuation without looking at the spatial distribution, in contrast to the hyperuniformity, which characterizes the long-range density fluctuation.

Here, we study how this local variance changes with $\mu$ and $\l$.
Figure \ref{fig:n2} shows the results for $\l=1$, 2 and 3.
An overall trend is that the local variance is maximized around $\mu=0$ and decreases as $\mu$ approaches $\pm\Emax$, as anticipated.
However, for $\l=1$, the local variance shows a dip around $\mu=0$, making a local minimum at $\mu=0$.
While the local variance increases monotonically with $\mu<0$ for $\l=2$, it shows a nonmonotonic dependence on $\mu<0$ for $\l=3$.
The difference between $\l\geq2$ and $\l< 2$ may be attributed to the presence/absence of the jump in the $n_i$ histogram.

\bibliography{ref}
\end{document}